\documentstyle[psfig,aps,twocolumn,prb]{revtex}
\begin{document}
%\wideabs{
\title{Degradation of LaMnO$_{3-y}$ surface layer in LaMnO$_{3-y}$/
metal interface} \author{A.Plecenik and K. Fr\"{o}hlich}
\address{Institute of Electrical Engineering, Slovak Academy of
Sciences, Dubravska cesta 9, 84239 Bratislava, Slovak Republic}
\author{J.P. Espin\'{o}s and J.P.Holgado}
\address{Instituto de Ciencia de Materiales de Sevilla,
Centro de Investigaciones Científicas "Isla de la Cartuja", C/ Americo Vespucio, s/n,
41092 Isla de la Cartuja, Sevilla, Spain}
\author{A. Halabica and M. Pripko}
\address{Institute of Electrical Engineering, Slovak Academy of Sciences,
Dubravska cesta 9, 84239 Bratislava, Slovak Republic }
\author{A. Gilabert}
\address{University of Nice Sophia Antipolis,  UMR CNRS 6622, Parc
Valrose, 06108 Nice Cedex 2, France}

\maketitle
\date{\today }

%}

%\narrowtext
\begin{abstract}

We report electrical measurements showing the degradation
processes of LaMnO$_{3-y}$ (LaMnO) in LaMnO/normal metal
interface in both point contact and planar-type junctions. Immediately
after the preparation of the interface, the degradation process was
followed by measuring the evolution of the junction resistance
versus time. This process is characterized by  the appearance of a
second maximum in the resistance vs. temperature (R-T) dependence at
temperatures lower than  the Curie temperature T$_c$, at which the
metal-insulator transition occurs in the bulk. These effects are
explained in terms of the formation of a depleted interface layer in
LaMnO caused by an out-diffusion of oxygen from the manganite
surface to the normal metal. This assumption is confirmed by XPS
measurement. Similar results on LaSrMnO$_{3-y}$ interfaces are
also obtained.

\end{abstract}

pacs: {    73.40.Cg, 68.35.Fx, 79.60.Jv}
%}

%\narrowtext

 The magnetoresistive rare earth perovskites La$_{1-x}$A$_x$MnO$_{3-y}$
(LaAMnO, A=Ca, Sr, Pb), which include LaMnO, have been in
the research spotlight for a decade. Optimally doped LaAMnO
exhibits a paramagnetic  - insulating behavior above the Curie
temperature T$_c$ and a ferromagnetic - metallic behavior below Tc. In
the recent years, the giant magnetoresistance phenomenon (MR) has been
intensively studied in these materials. However till low-field, MR
effects in granular films and tunneling of spin-polarized
electrons through an insulating barrier in LaAMnO/Insulator/LaAMnO
magnetic tunnel junctions (MTJ) gave promise to their
technological applications like magnetic field sensor and
nonvolatile magnetic random access memory.  One can expect a near
100\% value of tunneling magnetoresistance (TMR) in tunnel junctions
based on half metallic LaAMnO ferromagnet. TMR depends on the
quality of the magnetic state of the LaAMnO surface, interface
quality, surface roughness of the electrodes, etc. Anomalous
tunneling behavior observed in MTJ  could be explained by
extrinsic factors playing a dominant role in the MTJ interfaces.
This assumption has been made by several authors. Park et al.
\cite{Park98} showed that even in a fully oxidized LaSrMnO sample,
the magnetic properties at the surface boundary are significantly different
from those of the bulk. In addition, several experiments indicated
the creation of an oxygen depleted layer of LaAMnO  interfaces.
Mieville et al. \cite{Mieville98} studied the interface resistance
and transport across conducting ferromagnetic oxide/metal
interfaces. They measured a very high resistance of LaSrMnO/Al(Nb)
junctions, which could be  explained by the existence of a degraded
surface layer of ferromagnetic oxide due to a loss of oxygen. Also
de Teresa et al. \cite{Teresa00} observed a dependence of TMR
on the type of tunneling barrier. The present work was stimulated by
the above mentioned results, but mainly by the results
published in our previous paper \cite{Gilabert01}.
LaMnO/Al$_2$O$_3$/Nb junctions were fabricated for the study of
the influence of illumination on the electrical properties of MTJ.
The Al$_2$O$_3$ insulating barrier was prepared using well-known
thin films  Nb technology \cite{Gurwitch83}: An aluminium thin film was sputtered on LaMnO,
and then oxidized in an oxygen atmosphere. Finally a semi-transparent
Nb upper electrode was sputtered on the top. Using the
same technique with the same preparation parameters, the width of
the barrier created on the LaMnO was about two times higher than that
for Nb/Al$_2$O$_3$/Nb junctions.

In this paper we study the LaMnO
surface properties  in contact with Al, In, Au and Pb metals in
order to explain the physical processes in the interface which
give rise to an unusual behavior of MTJ. The
LaMnO/normal metal point contact exhibits a change of resistance immediately
after its preparation. In the planar junctions (more stable than point contact)
the appearance
of a  second maximum in the R-T dependence was observed several hours after junction preparation.
These effects are explained
by an out-diffusion of oxygen from LaMnO, and they were confirmed by XPS
spectroscopy.

200 nm thick LaMnO epitaxial thin films with T$_c$=270 K
were deposited on single crystalline SrTiO$_3$
substrates by low pressure liquid source metal-organic chemical
vapour  deposition (MOCVD) \cite{Frohlich95}. LaMnO/metal point
contacts were realized using a holder which exerted a constant
pressure of the tip onto the sample. Bulk Au, Al, or  Pb were used
for the upper electrode. The shape of the sharp point was prepared by
mechanical sharpening.  After the electrical measurements, a
200x200 $\mu m^2$ geometric area  was evaluated for the Pb tip.

For the preparation of planar junctions,
a 40 $\mu$m - wide and 260 $\mu$m - long LaMnO$_3$ base electrode
was formed by wet
etching through resist mask. After stripping the photoresist, the
area for deposition of next layers by lift-off technique was
defined in positive photoresist. Shortly before the deposition of
the barrier and the  upper electrodes, the surface of LaMnO$_3$
was etched to eliminate a contaminated and degraded upper layer. A
100 nm thick Al or In layer was deposited by thermal evaporation.
Junctions with an area of 40$\times$40 $\mu$m$^2$  were finalized
after the removal of the photoresist.\\ The R-T and resistance vs. time
(R-t) characteristics were measured by a computer controlled four
point method .

XPS spectra were obtained in an ESCALAB 210 spectrometer
that consisted of two separate independently-pumped chambers.
Pressures in the range of 6x10$^{-11}$ mbar and
10$^{-8}$ mbar were obtained in the analysis and preparation
chambers, respectively. Al has been evaporated in the analysis
chamber from a resistively heated filament made of Al wire wrapped
around a thick tungsten wire.  The power was maintained to
give a reproducible evaporation rate of 1 monolayer of Al per minute.
Control evaporation carried out onto a clean Au foil showed
that under our experimental conditions, only metallic Al is
evaporated, maintaining this oxidation state throughout all the
acquisition time.
 A hemispherical electron energy analyser working in the pass energy
constant mode at a value of 50 eV was used. Unmonochromatized Al
Ka radiation was used as the excitation source. Spectra were energy-calibrated
by taking the La3d5/2 peak at 834.6 eV (BE). The spectra were
acquired at 90 (normal) and 20 (grazing) degrees with respect to
sample surface.

In Fig 1, the LaMnO/metal point contact resistance dependence
versus time shows a significant increase at several hundred
seconds after the preparation of the contact. These changes were
observed even if noble metal (Au) was used as the upper electrode
material. Similar results on high-T$_c$ superconductors (HTS) were
described in our previous work \cite{Grajcar92}. The change of the
point contact resistance was explained within an oxygen
out-diffusion model.\\ The total point contact resistance with a
tip made of nonreactive metal (e.g. Au) can be expressed as $R =
R_M + R_{LaMnO} + R_T$, where $R_T$ is the tunneling resistance
and $R_M$ and $R_{LaMnO}$ the contact resistances with  metal and
LaMnO, respectively. We suppose that $R_M$ and $R_T$ are constant
in time for stable nonreactive metals, and the change of
resistance $R$ is then given by the change of $R_{LaMnO}$. The
creation of additional barriers from a LaMnO oxygen depleted layer
as well as from the oxide of the upper metallic electrode (created
from reactive metal like Al, Pb) creates a complex interface. But
in all cases, the Curie temperature $T_c$ in the LaMnO oxygen
depleted region must be changed also.

In the next measurements the planar junctions were used because they were more stable
than the point contact ones. Because Al is a very reactive metal with a very short
time of interaction, indium as an upper electrode was used for the following measurement.
The R-T characteristic measured on a LaMnO/In planar junction at
different times ( 3, 5 and 23 hours) after preparation of the contacts is
shown in Fig.2.  We observed  the classical maximum correlated
with the I-M transition of the bulk material at $T_c$ around 275 K of the bulk
material. This maximum did not change very much with time. It
means that the oxygen content in the bulk material does not change.
After a long time (23 hours)  a second maximum appears around
T'$_c$= 185 K in the R-T characteristics. We know that in
LaAMnO systems the Tc decreases with a decrease of the oxygen content,
and it disappears for deoxygenated samples that exhibit semiconducting
properties. This new maximum is related to  the oxygen
depleted interface, and it is broader due to a distribution
of T'$_c$. This value of T'$_c$= 185 K
corresponds to an average oxygen content at the interface of y
around 0.1.\cite{Leon-Guevara97}

To confirm the occurrence of out-diffusion processes in the LaMnO surface layer,
XPS measurements of LaMnO/Al interfaces were studied. Fig.3
showns spectra of the O1s, Mn2p, Al2p and La3d
regions measured on clean (original) LaMnO surface (curves a) at the normal
acquisition angle,
after the
deposition of around 1 monolayer of Al (curves b), at the normal adquisition
angle, around 2 monolayers of Al, at the normal adquisition angle (curves c),
and curves d were measured as curves c at a grazing adquisition angle.
One can see the following features in the XPS spectra after the deposition of
Al:\\ -   two peaks correlated with Al and Al$_2$O$_3$,\\ -   two peaks
in the O1s spectra correlated with oxygen bonding in LaMnO$_3$ at 529.9 eV and Al$_2$O$_3$
at 531.6 eV,\\ -   shift of the peaks in the Mn2p spectra
due to the decreasing of Mn valency.

The deposition of the first monolayer of Al on the LaMnO surface provokes the oxidation
of Al to Al$^{3+}$, whose peak arises at about 75 eV (BE). This behaviour was tested
by a simultaneous deposition of Al on Au. The detection of metallic Al during the
whole experiment (more than 1 hour) exclude the oxidation of Al due to the
residual O$_2$ partial preasure in the analytic chamber. In additional, when a second
amount of Al was deposited (curve c in Fig.3a), most of it remained
metallic (peak at 72.6 eV BE).
Finally, when this situation is examined at the grazing adquisition angle (curve d in Fig.3a),
the relative intensity of the peak from Al(0), in comparison with that from Al$^{+3}$, increases
substantially, indicating that metallic Al is located on the top of the Al$_2$O$_3$.

The oxidation of Al in the LaMnO$_3$/Al interface accompanying the appearence
of two peaks in the O1s spectra
is correlated with a change of binding energy due to the creation of
Al$_2$O$_3$. The smooth second peak was measured also on clean LaMnO surface (curve a in Fig.3c)
as well as after cleaning treatment. The origin of the second peak can be explained by the contamination of
LaMnO by CO$_3^=$ species within the preparation procedure of the LaMnO thin film.
Anyway, the intensity of the peak at the energy of 520.6 eV increase after the deposition
of Al (curves b and c in Fig.3c) due to the arising of O$^=$ ions in Al$_2$O$_3$. The measurement
at the grazing angle shows that Al$_2$O$_3$ is created on the LaMnO surface (the global intensity
of Al2p signal grows and the Mn and La signals decrease). These facts aslo indicate the
existence of metallic Al on the top of Al$_2$O$_3$.

A significant chemical
shift of Mn2p peak from 642.5 eV before Al evaporation to 641.4 eV
after Al evaporation strongly indicates a decrease of Mn valency to +3 state
with a subsequent change of the
Mn$^{3+}$/Mn$^{4+}$ ratio, typical for oxygen-deficient LaAMnO.

From XPS measurements described above, we can confirm the loss of
oxygen from the LaMnO in the LaMnO/Al interface and the creation of
a LaMnO/Al$_2$O$_3$/Al junction with a subsequent change in the
Mn$^{3+}$/Mn$^{4+}$ ratio.

In conclusion, we presented R-T and R-t measurements  on
LaMnO/metal point contact and planar junctions. The time evolution
of point contact junction resistance as well an arising of a second
peak on the R-T characteristics were observed. These effects were
explained in terms of the formation of a depleted interface layer in
LaMnO$_{3-x}$ caused by the out-diffusion of oxygen from the manganite
surface to the normal metal. This assumption was confirmed by XPS
measurements .
The similar
results (XPS as well as time evolution of point contact
resistance) on LaSrMnO magnetoresistive thin films were also obtained.

\section*{Acknowledgement}
This work
was supported by the Slovak Grant Agency for Science (Grants
No.2/7199/20 and 1/7072/20) and in part by the European Commission
(project GRT-CT-2000-05001 MULTIMETOX)

\begin{figure}
\centerline{ \psfig{file=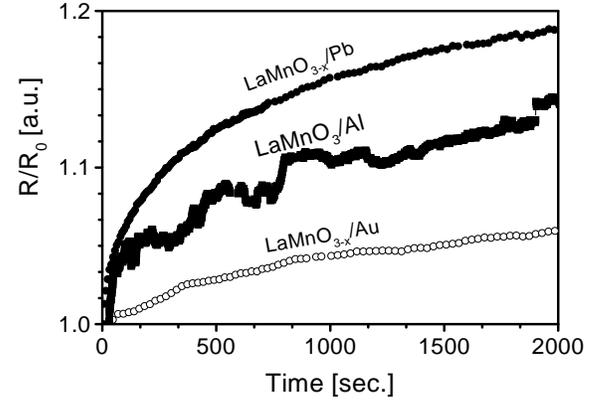,width=\columnwidth,angle=0} }
\caption{ Time evolution of the LaMnO/metal (Au - open circles, Al
- solid squares and Pb - solid circles) point contact resistance.
The measurement was started immediately after the point
contact preparation.} \label{fig:Rt}
\end{figure}

\begin{figure}
\centerline{ \psfig{file=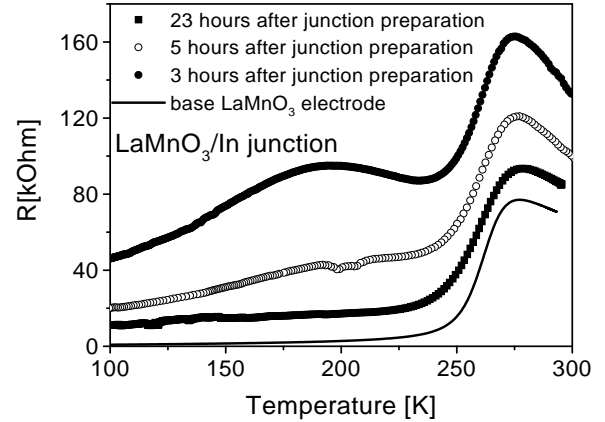,width=\columnwidth,angle=0} }
\caption{R-T dependence of the LaMnO electrode (solid
line) and R$_T$-T dependence of LaMnO/In planar junction measured
3 hours (solid sguares), 5 hours (open
circles) and 23 hours (solid circles) after the deposition of the In upper electrode. }
\label{fig:RT}
\end{figure}

\begin{figure}
\centerline{ \psfig{file=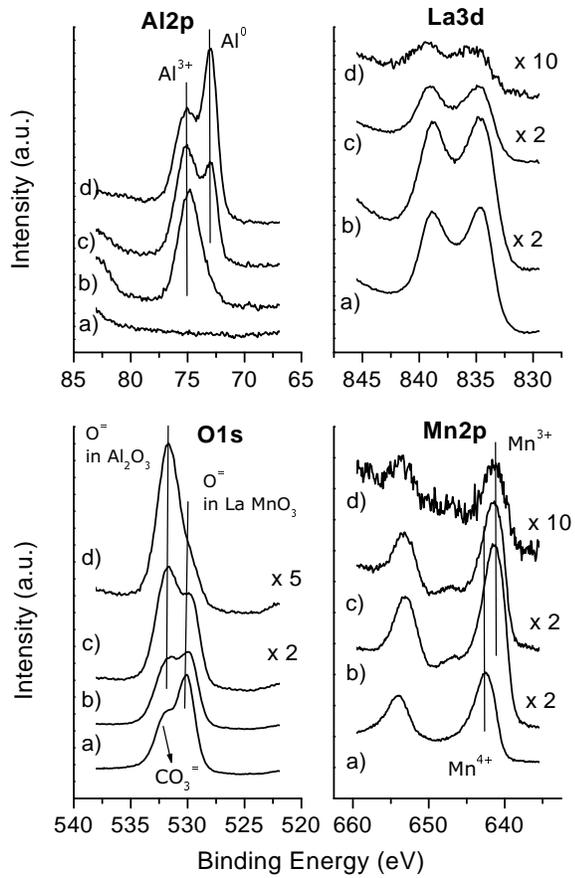,width=\columnwidth,angle=0} }
\caption{ \\XPS spectra of the O1s, Mn2p, Al2p and
La3d regions measured on the clean surface of LaMnO (curves a)
after the deposition of one monolayer and two monolayers of Al, at the normal adquisition angle (curves
b and c, respectively) and the same that curves c, at the grazing adquisition angle.} \label{fig:XPS}
\end{figure}


\begin{references}

\bibitem{Park98}J.-H.Park, E.Vescovo. H.-J.Kim, C.Kwon, R.Ramesh
and T.Venkatesan {\bf 81}, Phys.Rev.Lett. {\bf 81}, 1953 (1998).
\bibitem{Mieville98} L.Mieville, D.Worledge, T.H.Geballe, R.Contreras and
K.Char, Appl.Phys.Lett. {\bf 73}, 1736 (1998).
\bibitem{Teresa00} J.M. de Teresa, A.Barth\'el\'emy, J.P.Contour
and A.Fert, J.Magn.Magn.Mater. {\bf 211}, 160 (2000).
\bibitem{Gilabert01}A.Gilabert, A.Plecenik, K.Fr\"{o}hlich,
\v{S}.Ga\v{z}i, M.Pripko, \v{Z}.Mozolov\'{a}, D.Machajd\'{\i}k,
\v{S}.Be\v{n}a\v{c}ka and M.G.Medici, Appl.Phys.Lett. {\bf 78},
1712 (2001).
\bibitem{Gurwitch83} M.Gurwitch, M.A.Washington and H.A.Huggins, Appl.Phys.Lett.
{\bf 42}, 472 (1983).
\bibitem{Frohlich95} K. Fr\"{o}hlich, J. \v{S}ouc, D. Machajd\'{\i}k, A.P.
Kobzev, F. Weiss, J. P. Senateur, K.H. Dahmen, Journ. de Physique
{\bf IV C5}, 533 (1995).
\bibitem{Grajcar92}M.Grajcar, A.Plecenik, \v{S}.Be\v{n}a\v{c}ka,
Ju.Revenko, V.M.Svistunov, Physica C {\bf 218}, 82 (1993).
\bibitem{Leon-Guevara97} A.M.De Leon-Guevara; P.Berthet; J.Berthon;
F.Millot; R.Revcolevschi; A.Anane; C.Dupas; K.Le Dang; J.P.Renard; P.Veillet
Phys.Rev. B {\bf56}, 6031(1997)
\end{references}
\end{document}